\documentclass[runningheads]{llncs}
\usepackage[T1]{fontenc}
\usepackage{graphicx}
\usepackage{amsmath}  
\usepackage{float}
\usepackage{tabularx}
\usepackage{colortbl}
\usepackage[table]{xcolor}
\usepackage{caption}
\usepackage{subcaption}
\usepackage[table]{xcolor}
\definecolor{rowgray}{gray}{0.9}    
\definecolor{rowwhite}{gray}{1.0}
\definecolor{green}{RGB}{162, 202, 154}
\definecolor{orange}{RGB}{243, 208, 184}
\definecolor{red}{RGB}{255, 130, 130}
\usepackage{booktabs}
\usepackage{arydshln}
\usepackage{array}
\newcolumntype{C}[1]{>{\centering\arraybackslash}p{#1}}  
\usepackage{orcidlink}
\usepackage{url}
\usepackage{hyperref}

\begin{document}
\title{Towards Early Detection: AI-Based Five-Year Forecasting of Breast
Cancer Risk Using Digital Breast Tomosynthesis Imaging}
\titlerunning{AI-Based Five-Year Forecasting of Breast Cancer Using DBT Imaging}

\author{Manon A. Dorster \inst{1,2}\orcidlink{0009-0008-2679-6218} \and
Felix J. Dorfner \inst{1, 3}\orcidlink{0009-0003-5914-3233} \and
Mason C. Cleveland \inst{1}\orcidlink{0009-0003-4472-043X} \and
Melisa S. Guelen \inst{1, 3}\orcidlink{0009-0001-2187-0895} \and
Jay Patel \inst{1}\orcidlink{0000-0002-8507-795X} \and 
Dania Daye \inst{1,4} \and
Jean-Philippe Thiran \inst{2} \and
Albert E. Kim \inst{1}\orcidlink{0000-0003-2359-6209} \and
Christopher P. Bridge \inst{1}\orcidlink{0000-0002-2242-351X}}

\authorrunning{Manon A. Dorster}
%

\institute{Athinoula A. Martinos Center for Biomedical Imaging, Mass General Brigham and Harvard Medical School, Charlestown, MA, USA \newline
\email{akim46@mgh.harvard.edu}, \email{cbridge@mgh.harvard.edu} \and
École Polytechnique Fédérale de Lausanne (EPFL), Lausanne, Switzerland \and {Charité – Universitätsmedizin Berlin, corporate member of Freie Universität Berlin and Humboldt-Universität zu Berlin, Berlin, Germany} \and {Department of Radiology, University of Wisconsin School of Medicine and Public Health, Madison, WI, USA} 
\\
}

\maketitle              

\begin{abstract}
As early detection of breast cancer strongly favors successful therapeutic outcomes, there is major commercial interest in optimizing breast cancer screening. However, current risk prediction models achieve modest performance and do not incorporate digital breast tomosynthesis (DBT) imaging, which was FDA-approved for breast cancer screening in 2011.  To address this unmet need, we present a deep learning (DL)-based framework capable of forecasting an individual patient's 5-year breast cancer risk directly from screening DBT. Using an unparalleled dataset of 161,753 DBT examinations from 50,590 patients, we trained a risk predictor based on features extracted using the Meta AI DINOv2 image encoder, combined with a cumulative hazard layer, to assess a patient's likelihood of developing breast cancer over five years. On a held-out test set, our best-performing model achieved an AUROC of 0.80 on predictions within 5 years. These findings reveal the high potential of DBT-based DL approaches to complement traditional risk assessment tools, and serve as a promising basis for additional investigation to validate and enhance our work. 

\end{abstract}
\section{Introduction}

Breast cancer (BC) is the leading cause of cancer-related mortality in women worldwide \cite{who}. Given recent advances in targeted and endocrine therapies, early diagnosis of breast cancer is strongly linked to higher curative potential in women affected by the disease. Consequently, there is a major interest in developing optimal screening tools that can both maximize early detection and reduce false-positive assessments. Current risk assessment models (e.g., Tyrer-Cuzick model) \cite{tyrer_cuzick} evaluate known risk factors, such as family history and breast density, but achieve modest accuracy. To this end, computational models employing 2D mammography, which contain orthogonal data compared to conventional risk factors, have demonstrated superior performance over traditional risk assessment models in predicting future breast cancer risk \cite{Yala}. Therefore, there is considerable interest in optimizing end-to-end DL models, with breast imaging data given as input, to better forecast future breast cancer risk. 

Digital breast tomosynthesis (DBT) was FDA approved for breast cancer screening in 2011, and offers advantages over 2D mammography.  2D mammography is limited in its ability to reveal small tumors due to superimposed breast tissues that may mask abnormalities, while DBT imaging offers near-3D imaging capabilities \cite{dbt_vs_mammo}. Current work in the field of AI prediction models based on breast imaging mainly applies to 2D mammography data \cite{Yala}\cite{wang}. Further studies leverage mammography imaging data from multiple time points, thus exploiting longitudinal variability of image features \cite{karaman}\cite{wang2}\cite{lee}\cite{dadsetan}. A study led by Eriksson et al. \cite{eriksson} explores one-year breast cancer risk prediction based on DBT input data. A five-year time frame is recommended for breast cancer risk prediction \cite{nccn}; however, no long-term risk prediction model based on DBT data currently exists. Therefore, using an unparalleled dataset of DBTs from Mass General Brigham, we present a DL-based prototype of a 5-year risk assessment model, leveraging the pre-trained Meta AI DINOv2 \cite{dino} image feature encoder followed by a custom fully-connected layer. 


\section{Methods}

\subsection{Dataset} \label{sec:dataset}

For model training, development, and testing, we compiled a dataset of 161,753 screening DBT examinations from 50,590 patients at Mass General Brigham. All DBTs included in the dataset are \textbf{screening} DBTs conducted on female patients between January 2012 and November 2023. Moreover, the dataset is divided into two cohorts: the \textbf{pre-cancer} group, consisting of 6,449 DBT examinations from 2,915 patients prior to biopsy-confirmed breast cancer, and the \textbf{healthy} group, which comprises of 155,304 DBT examinations from 47,675 patients unaffected by breast cancer. As most patients received multiple DBT examinations, we defined a data point in our dataset as a \textbf{DBT study} conducted for a given patient at a given time point. The \textbf{study date} or time $\mathbf{t_0}$ of a DBT study is the time at which the examination was conducted. For the pre-cancer cohort, the \textbf{diagnosis date} is the date at which the patient was first diagnosed with biopsy-confirmed  breast cancer. Year 1 spans from the study date ($t_0$) to 365 days later, year 2 from $t_1$=$t_0$+365 days to 365 days after $t_1$, etc.

A positive case is defined as a DBT study conducted within five years prior to a breast cancer diagnosis. Diagnosis dates in the pre-cancer cohort were extracted using the Qwen 2.5 72B large language model \cite{qwen} on all available breast pathology reports. Our implementation excludes DBT studies conducted within less than 6 months of breast cancer diagnosis from the validation and test sets. 
For the healthy cohort, since some DBT studies are recent and lack full 5-year follow-up, we retained only negative cases with at least one year of screening follow-up. Combined with the label masking process (cf. Section \ref{sec:labeling}), this ensures that the healthy dataset includes only confirmed negatives, excluding DBT studies from patients lost to follow-up or who may develop cancer in the future.

Only studies with available right and left cranio-caudal and medio-lateral oblique views (RCC, LCC, RMLO, and LMLO) were included. Each view acquired during a DBT study is saved as an individual DICOM file, which we refer to as a \textbf{DBT series} in the rest of this work. The dataset was divided into training, validation, and test subsets, corresponding to 70\%, 15\% and 15\% of the data, respectively. All studies for a unique patient were assigned to the same subset. Due to the high imbalance between the pre-cancer and healthy groups, which reflects natural breast cancer occurrence among women, we used a weighted sampler in order to load the data from the two groups in equal proportions during training. 

\subsubsection{Labeling and Masking} \label{sec:labeling}

The output of the risk prediction model is a 5-element vector, each element corresponding to the predicted \textbf{breast cancer status} for years 1 through 5. Labeling and masking are performed using a \textbf{sliding window} approach, similar to the framework in \cite{Yala}, such that each DBT study for a given patient receives its own label and mask corresponding to the five-year period after acquisition of the study. Hence, the risk assessment model is robust to variable amounts of follow-up. A label at year $n$ indicates whether the patient had a positive (1) or negative (0) cancer status at that specific time point, as shown in Eq. \ref{eq:binary_labels}. If breast cancer is diagnosed at year $n$, the label is also marked positive for years $n+1$ to 5.
\begin{equation} \label{eq:binary_labels}
\mathbf{y} = [y_1, ..., y_5], \quad \text{with } y_i =
\begin{cases}
0, & \text{if negative at year } i, \\
1, & \text{if positive at year } i,
\end{cases}
\quad \text{for } i \in \{1, ..., 5\}
\end{equation}
Variability in the amount of follow-up among patients in the dataset calls for a masking strategy. We defined a mask vector \textbf{w} for each study with five binary entries, each indicating whether breast cancer status is available for the corresponding year.  Years with no information regarding the malignancy status
of a given patient were masked out during training, validation, and inference.

\subsubsection{Dataset Statistics}
Patients in the dataset display age ranges of 68.8±11.7 years and 65.1±10.9 years for the pre-cancer and healthy cohorts, respectively. All DBT studies were acquired using Hologic Selenia Dimensions systems. The demographics of the dataset are detailed in Table \ref{tab:demographics}.

\begin{table}[tb]
    \centering
    \begin{tabular}{p{6cm}C{3cm}C{3cm}}
        \centering
         & & \\
        \hline
        \rowcolor{gray!20}
        \textbf{Breast Density (Patient Counts)} & \textbf{Healthy} & \textbf{Pre-cancer}\\
        \hline
        a - almost entirely fatty & 4,785 (10\%) & 117 (4\%) \\ 
        \hdashline
        b - scattered areas of fibroglandular density & 22,929 (48\%) & 1,424 (49\%) \\
        \hdashline
        c - heterogeneously dense & 17,371 (37\%) & 1,060 (36\%) \\
        \hdashline
        d - extremely dense & 2,537 (5\%) & 104 (4\%) \\
        \hdashline
        Unknown & 53 ($<$1\%) & 210 (7\%) \\
        \hline
        \rowcolor{gray!20}
        \textbf{Ethnicity (Patient Counts)} & \textbf{Healthy} & \textbf{Pre-cancer} \\
        \hline
        White & 38,458 (81\%) & 2,467 (85\%) \\
        \hdashline
        Black & 2,762 (6\%) & 129 (4\%) \\
        \hdashline
        Asian & 2,156 (4\%) & 117 (4\%) \\
        \hdashline
        Other/Unknown & 4,299 (9\%) & 202 (7\%) \\
        \hline
    \end{tabular}
    \begin{tabular}{p{6cm}C{2cm}C{2cm}C{1.9cm}}
        \rowcolor{gray!20}
        \textbf{BC Subtypes (Patient Counts)} & \textbf{Train} & \textbf{Validation} & \textbf{Test} \\
        \hline
        Invasive ductal carcinoma (IDC) & 1,691 (69\%) & 158 (69\%) & 166 (71\%) \\
        \hdashline
        Invasive lobular carcinoma (ILC) & 289 (12\%) & 30 (13\%) & 28 (12\%) \\
        \hdashline
        IDC \& ILC & 137 (5\%) & 11 (5\%) & 7 (3\%) \\
        \hdashline
        Ductal carcinoma in situ (DCIS) & 186 (8\%) & 16 (7\%) & 17 (7\%) \\
        \hdashline
        Other & 148 (6\%) & 14 (6\%) & 17 (7\%) \\
        \hline
        \rowcolor{gray!20}
        \textbf{Years to Cancer (Study Counts)} & \textbf{Train} & \textbf{Validation} & \textbf{Test} \\
        \hline
        1 year & 1,718 (31\%) & 18 (4\%) & 24 (5\%) \\
        \hdashline
        2 years & 1,182 (22\%) & 132 (28\%) & 154 (31\%) \\
        \hdashline
        3 years & 999 (18\%) & 126 (27\%) & 121 (24\%) \\
        \hdashline
        4 years & 813 (15\%) & 103 (22\%) & 104 (21\%) \\
        \hdashline
        5 years & 774 (14\%) & 86 (19\%) & 95 (19\%) \\
        \hline
        \rowcolor{gray!20}
        \textbf{BI-RADS (Study Counts)} & \textbf{Training} & \textbf{Validation} & \textbf{Test} \\
        \hline
        0 - need for additional imaging & 7,790 (6\%) & 1,206 (6\%) & 1,157 (5\%) \\
        \hdashline
        1 - negative & 93,495 (80\%) & 17,646 (80\%) & 17,477 (81\%) \\
        \hdashline
        2 - benign & 16,899 (14\%) & 3,092 (14\%) & 2,958 (14\%) \\
        \hdashline
        3 - probably benign & 15 ($<$1\%) & 2 ($<$1\%) & 7 ($<$1\%) \\
        \hdashline
        $\geq$4 - suspicious, suggestive of malignancy & 9 ($<$1\%) & 0 (0\%) & 0 (0\%) \\
        \hline
    \end{tabular}
    \captionsetup{font=small}
    \caption{Demographics of the full dataset. Percentages are relative to each column subset. For ``years to cancer'', percentages are within pre-cancer cases only.}
    \label{tab:demographics}
\end{table}

Breast radiology reports assign scores using the Breast Imaging Reporting and Data System (BI-RADS) \cite{birads}, which also classifies breast density into four categories \cite{density_class} (cf. Table \ref{tab:demographics}). High density has been shown to correlate with increased breast cancer risk \cite{rsna}. A few density labels in our dataset are missing (cf. Table \ref{tab:demographics}), thus we excluded those cases from the density subgroup analysis (Table \ref{tab:subgroup_analysis}). Screening cases are rarely attributed a BI-RADS score above 3 (see Table \ref{tab:demographics}); suspicious findings are typically labeled BI-RADS 0 and are referred for diagnostic imaging or biopsy.

\subsection{Risk Assessment Model} \label{sec:model}

\begin{figure}[tb]
    \centering
    \includegraphics[width=\linewidth]{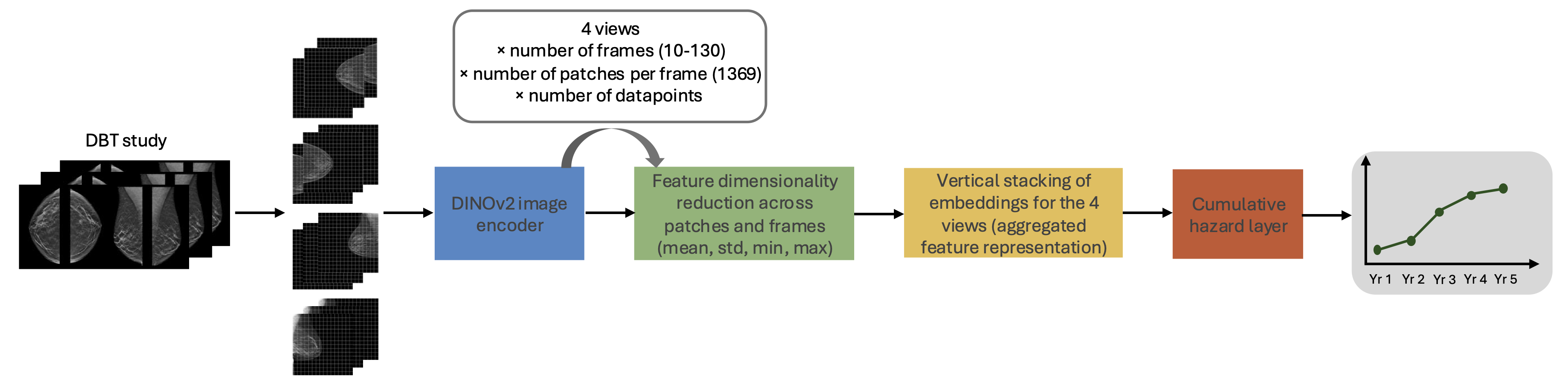}
    \caption{5-year risk assessment model pipeline, inspired from Fig. 1 in \cite{Yala}.}
    \label{fig:model_archi}
\end{figure}

Image feature representations were obtained using the pre-trained DINOv2 foundation model (see Fig. \ref{fig:model_archi}). Using this state-of-the-art image encoder allowed for rapid model prototyping and reduced the substantial computational resources needed for development of a task-specific model. Embeddings were saved to disk and used for experiments in which we fed the feature representations to a linear layer, termed the \textbf{cumulative hazard layer}, which outputs one risk score per year following the study date.

\subsubsection{Image Encoding and Feature Extraction}

Each DBT study is composed of four views, with each view (DBT series) containing multiple 2D slices (\textbf{frames}). For every frame of a DBT series, we resized each frame to 518 by 518 pixels and ran a forward pass through the DINOv2 ViT-B/14 image encoder, from which we omitted the classifier head. Each forward pass yields one CLS token and 1369 patch token embeddings per frame, all of size 768. Embeddings for all frames in a DBT series are then stacked to obtain a feature representation of the full DBT series. Given that there are four DBT series per DBT study (RCC, LCC, RMLO, and LMLO), this results in a substantial number of features per study. To reduce dimensionality while attempting to preserve informative content, we computed summary statistics — mean, standard deviation (SD), minimum, and maximum — of each feature across patches within each frame and across frames within each DBT series when using the patch token representations. Following standard vision transformer architectures, one CLS embedding, typically used for downstream image-level classification tasks, is extracted for each 2D frame and captures global characteristics of the image. By contrast, patch embeddings capture patch-level representations. For each DBT series, we computed the summary statistics across frames. Different combinations of patch or CLS summary statistic embeddings were used for the downstream training of the cumulative hazard layer, as displayed in Table \ref{tab:auroc_test}.

\subsubsection{Cumulative Hazard Layer} \label{sec:hazard_layer}

Similarly to the implementation in \cite{Yala}, the cumulative hazard layer is a fully connected linear layer that takes as input the aggregated representation of a DBT study, consisting of the stacked feature representations of all four DBT series in the study. The cumulative hazard layer then outputs five predicted risk probabilities for each of the corresponding years following the study date.

We define ground truth labels \textbf{y} as described in Eq. \ref{eq:binary_labels}. The risk probabilities over 5 years should follow a monotonically increasing trend. It would not be coherent for a predicted risk at year $n$ to be higher than that at year $n+1$, hence the use of a cumulative risk probability framework. A fully-connected linear layer is applied to the feature representation of a DBT study and yields five output values. These logit outputs are passed through the softplus activation function to ensure non-negative values, which are then cumulatively summed across the years, producing monotonically increasing risk scores. Since the softplus function outputs values in the range $[0, \infty)$, a final normalization step is applied to map the cumulative predictions into valid probability values within the range $[0, 1]$.
 
Given an aggregated feature representation of a DBT study $x$, the probability that the concerned patient will develop breast cancer $k$ years after the study date is given by the sum of the independent risk probabilities for years 1 through $k$. Eq. \ref{eq:risk_prob} summarizes the cumulative risk probability layer output at a given year $k$, $\hat{y_i}$ being the model risk prediction at year $i$:
\begin{equation} \label{eq:risk_prob}
    P(t_{cancer}=k|x) = p_k = 1-\exp(-\sum_{i=1}^{k}\hat{y_i}) , \quad 1 \leq k \leq 5
\end{equation}
In the model training experiments, we aimed to minimize the risk prediction objective defined as the sum of the binary cross-entropy (BCE) losses for each year weighted by the mask \textbf{w}, averaged over N samples in a batch (Eq. \ref{eq:bce}).
\begin{equation}
\mathcal{L}_{BCE} = \frac{1}{N} \sum_{i=1}^{N} \frac{\sum_{j=1}^{5} w_{i,j} \left[ y_{i,j} \log p_{i,j} + (1 - y_{i,j}) \log (1 - p_{i,j}) \right]}{\sum_{j=1}^{5} w_{i,j}}
\label{eq:bce}
\end{equation}

\section{Results and Discussion} \label{sec:results}

AUROC results for the different risk assessment models we developed are shown in Table \ref{tab:auroc_test}. We observed that models based on CLS embeddings, which learn a global representation of an image \cite{tokens}, displayed overfitting trends, with the validation loss measured during model training starting to increase after the first few epochs. Therefore, we find that using patch tokens allows for the model to better capture detailed features \cite{tokens} relating to local regions that are relevant to risk prediction (see Table \ref{tab:auroc_test}). Furthermore, better performance is achieved when considering the entire distribution over patch embeddings by calculating mean, standard deviation, minimum and maximum statistics over all patches (\textbf{model 3} defined in Table \ref{tab:auroc_test}) rather than solely the mean (model 1 defined in Table \ref{tab:auroc_test}). AUROC values based on density subgroups computed with model 3 predictions on test set data (Table \ref{tab:subgroup_analysis}) highlight the robustness of the model across clinically relevant subgroups.

\rowcolors{2}{gray!20}{white}
\begin{table}
\centering
\begin{tabularx}{\textwidth}{|p{1.2cm}|X|p{3cm}|p{1.3cm}|X|X|X|X|X|}
\hline
\rowcolor{gray!40}
\textbf{Model Name} & \textbf{Embed. Type} & \textbf{Summary Stats} & \textbf{Year 1*} & \textbf{Year 2} & \textbf{Year 3} & \textbf{Year 4} & \textbf{Year 5} \\ \hline
Model 1 & patch & mean & 0.6682 & 0.7082 & 0.7495 & 0.7775 & \cellcolor{orange}0.7941 \\ 
Model 2 & patch & mean, SD & \cellcolor{red}0.6663 & 0.7063 & \cellcolor{orange}0.7514 & 0.7776 & 0.7915 \\ 
Model 3 & patch & mean, SD, min, max & 0.6758 & \cellcolor{green}0.7175 & \cellcolor{green}0.7544 & \cellcolor{green}0.7886 & \cellcolor{green}0.7999 \\ 
Model 4 & CLS & mean & 0.6746 & 0.6993 & 0.7399 & \cellcolor{orange}0.7779 & 0.7908 \\ 
Model 5 & CLS & mean, SD & \cellcolor{green}0.6980 & \cellcolor{orange}0.7087 & 0.7363 & 0.7680 & 0.7856 \\ 
Model 6 & CLS & mean, SD, min, max & \cellcolor{orange}0.6923 & \cellcolor{red}0.6934 & \cellcolor{red}0.7291 & \cellcolor{red}0.7658 & \cellcolor{red}0.7837 \\
\hline
\end{tabularx}
\captionsetup{font=small}
\caption{Yearly AUROC metrics for risk assessment model predictions on the test set (498 positive, 21,101 negative DBT studies). Cell colors represent the best (green), second best (orange), and worst (red) AUROC for each year. *Metrics considered unreliable due to small sample (24 positives).}
\label{tab:auroc_test}
\end{table}

\begin{figure}[H]
    \centering
    \includegraphics[width=0.8\linewidth]{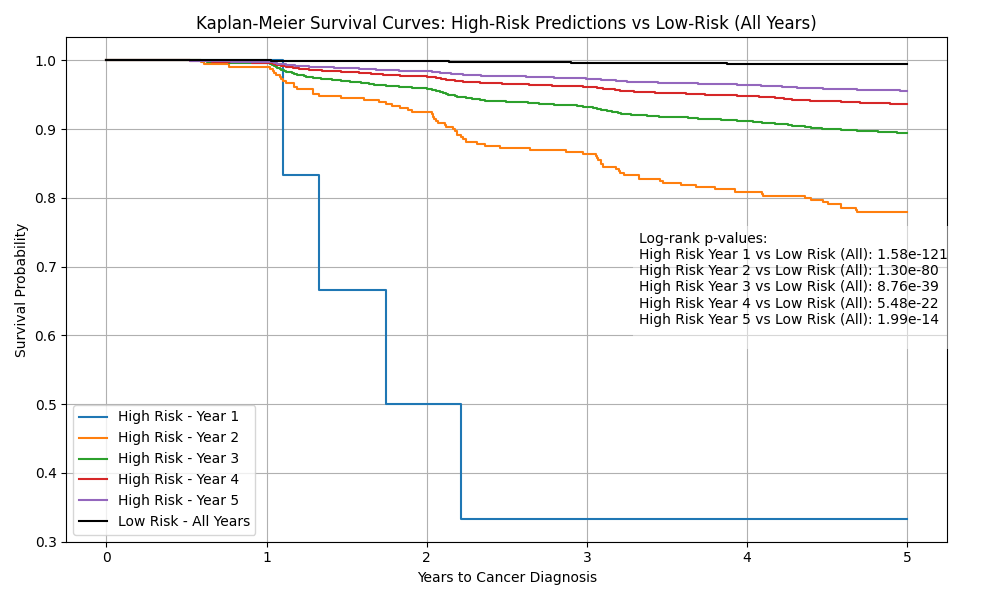}
    \captionsetup{font=small}
    \caption{Kaplan-Meier survival curves showing the estimated cancer-free survival probabilities over 5 years for different risk groups based on model 3 predictions on test set data. "Low Risk": breast cancer status predicted negative for all 5 years; "High Risk - Year $n$": breast cancer status predicted positive at year $n$. A binarization threshold for all years (0.3027) was computed by maximizing Youden's J-statistic.}
    \label{fig:kaplan-meier}
\end{figure}

The Kaplan-Meier curves in Figure \ref{fig:kaplan-meier} address whether cases predicted as high-risk at year $n$ develop cancer earlier than those predicted as low-risk at year $n$ or across all 5 years. These trends demonstrate a clear separation between cancer-free survival in each group and highlight the model's ability to efficiently stratify the DBT studies into risk groups.

To our knowledge, our dataset of 50,590 patients and 161,753 screening DBTs is one of the largest used to train a breast cancer risk prediction model using end-to-end DL. Our work demonstrates promise in applying DL to screening DBTs to quantify 5-year future risk.  Interestingly, we note that the performance of our model is comparable to state-of-the-art risk models \cite{Yala} employing 2D mammograms as input. Therefore, given the emerging availability of DBTs for breast cancer screening, further efforts are warranted to optimize DBT-based models for risk prediction.  

Leveraging the existing DINOv2 model allowed for rapid model development on this large dataset and achieved promising proof-of-concept results, but training a task-specific network (either a convolutional neural network (CNN) or vision transformer) may allow for better performance. Our efforts to train such a network were limited by the available computational resources.
Future work will focus on comparisons to trained, task-specific image encoders tailored for DBT imaging to evaluate the suitability of DINOv2 as an image encoder for our task. Moreover, as DINOv2 was pre-trained on natural 2D images and lacks 3D spatial awareness, it is not ideally suited to capture the pseudo-3D characteristics of DBT volumes. To address this limitation, we are currently exploring self-supervised learning methods leveraging large amounts of unlabeled DBT images to develop pre-training strategies capable of capturing the essence of DBT data. 

Another promising avenue would be more sophisticated feature dimensionality reduction methods that better characterize biologically relevant regions of DBT volumes (e.g., attention-based multiple instance learning for feature selection) \cite{mil}, rather than simply using the summary statistics of the patch or CLS embeddings. In addition, as multi-modal models have demonstrated improved performances compared to image-only models \cite{Yala}, we also aim to pursue our work by integrating both DBT data and clinical risk factors in our model, such as age and hormonal factors, similarly to the Mirai framework in \cite{Yala}. Finally, we note that our dataset is comprised predominantly of White breast cancer patients (see Table \ref{tab:demographics}). To this end, future studies should recruit a more ethnically diverse dataset in order to augment model generalizability and robustness.  

\renewcommand{\arraystretch}{1.2} 
\begin{table}[tb]
\centering
\begin{tabular}{|p{3.8cm}|p{2cm}|p{2cm}|p{2cm}|p{2cm}|p{2cm}|}
\hline
\rowcolor{gray!40}
\textbf{Breast Density} & \textbf{a*} & \textbf{b} & \textbf{c} & \textbf{d*} \\
\hline
\textbf{\# Patients Total} & 529 (10\%) & 2,657 (49\%) & 1,913 (35\%) & 301 (6\%) \\ 
\textbf{\# Pre-cancer Patients} & 6 (3\%) & 110 (51\%) & 82 (38\%) & 16 (8\%) \\
\textbf{\# Healthy Patients} & 523 (10\%) & 2,547 (49\%) & 1,831 (35\%) & 285 (6\%) \\
\textbf{\# DBT Studies} & 1,996 & 10,951 & 7,470 & 1,119 \\
\textbf{Year 1 AUROC} & 0.8417 & 0.4587 & 0.7177 & 0.8083 \\
\textbf{Year 2 AUROC} & 0.8853 & 0.7259 & 0.7200 & 0.6521 \\
\textbf{Year 3 AUROC} & 0.8486 & 0.7624 & 0.7513 & 0.6564 \\
\textbf{Year 4 AUROC} & 0.8373 & 0.8037 & 0.7731 & 0.6963 \\
\textbf{Year 5 AUROC} & 0.8668 & 0.8174 & 0.7762 & 0.7166 \\
\hline
\end{tabular}
\vspace{0.3cm}
\captionsetup{font=small}
\caption{Breast density subgroup analysis based on model 3 predictions on test set data. Density labels: a - almost entirely fatty, b - scattered areas of fibroglandular density, c - heterogeneously dense, d - extremely dense. *Small number of positive cases, to interpret with caution.}
\label{tab:subgroup_analysis}
\end{table}

\section{Conclusion}
Using an unparalleled dataset of 50,590 patients and 161,753 screening DBTs, our study presents one of the first efforts to develop a DL-based model capable of reliably predicting five-year breast cancer risk using DBT imaging. Our study demonstrates that DBT, a modality routinely used in clinical screening, contains predictive imaging biomarkers of future cancer development that could possibly outperform those derived from 2D mammograms. The model achieves promising performances across all five years, highlighting the potential of DBT-based risk models to support personalized screening strategies. Future work will focus on improving model performance through DBT-specific pre-training and incorporating multi-modal clinical information. Ultimately, this work represents a significant step toward the integration of AI-driven risk prediction into routine breast cancer screening, paving the way for optimized use of clinical resources aimed at reducing late-stage diagnoses and improving survival outcomes.

%
%
%
%
\bibliographystyle{splncs04}
\bibliography{references}

\end{document}